\documentclass[twocolumn,showpacs,preprintnumbers,amsmath,amssymb]{revtex4}

\usepackage{graphicx} % Include figure files
\usepackage{dcolumn} % Align table columns on decimal point
\usepackage{bm} % bold math

\begin{document}

\title{Discovery potential of the next-to-minimal supergravity-motivated model}

\author{Csaba Bal\'azs} %\email{First.Author@institution.edu}
\author{Daniel Carter} %\email{Second.Author@institution.edu}
\affiliation{School of Physics, Monash University, Melbourne Victoria 3800, Australia}

\date{August 31, 2008}

\begin{abstract}

Applying a likelihood analysis to the next-to-minimal supergravity-motivated 
model, we identify parameter space regions preferred by present experimental 
limits from collider, astrophysical, and low energy measurements. We then show 
that favored regions are amenable to detection by a combination of the CERN Large 
Hadron Collider and an upgraded Cryogenic Dark Matter Search, provided that the 
more than three sigma discrepancy in the difference of the experimental and the 
standard theoretical values of the anomalous magnetic moment of the muon 
prevails in the future.

\end{abstract}

% 12.60.Jv Supersymmetric models
% 14.80.Ly Supersymmetric partners of known particles  
% 95.35.+d Dark matter 
% 13.40.Em Electric and magnetic moments  
% 95.30.Cq Elementary particle processes   
 
  % PACS, the Physics and Astronomy
                                   % Classification Scheme.
 %Use showkeys class option if keyword display desired

\maketitle

\section{\label{sec:Introduction}Introduction}

% SUSY 

Supersymmetry is very successful in solving outstanding problems of the standard 
model (SM) of elementary particles.  The theory naturally explains the dynamics 
of electroweak symmetry breaking while preserving the hierarchy of fundamental 
energy scales, it incorporates dark matter and the asymmetry between baryons and 
antibaryons, it reconciles the unification of gauge forces and accommodates 
gravity, and more \footnote{See Ref.~\cite{Peskin:2008nw} and references therein.}. 
Thus, it is important to examine the prospects of the CERN Large Hadron Collider 
(LHC) finding supersymmetry.

% MSSM

One of the main motivations for supersymmetry is that it can naturally bridge 
the hierarchy between the weak and Planck scales.  Unfortunately, the presence 
of the superpotential $\mu$ term in the minimal supersymmetric extension of the 
standard model (MSSM) undermines this very aim \cite{Kim:1983dt}.  Experimental 
data have also squeezed the MSSM into fine-tuned regions, creating the 
supersymmetric little hierarchy problem \footnote{For a review see 
Ref.~\cite{Giudice:2008bi}.}.
%
% XMSSM
%
Extensions of the MSSM by gauge singlet superfields not only resolve the $\mu$ 
problem, but can also reduce the little hierarchy \cite{Dermisek:2005ar, 
BasteroGil:2000bw, Gunion:2008kp}.  In the next-to-minimal MSSM (NMSSM), the 
$\mu$ term is dynamically generated and no dimensionful parameters are 
introduced in the superpotential (other than the vacuum expectation values that 
are all naturally weak scale), making the NMSSM a truly natural model 
\cite{Fayet:1974pd, Nilles:1982dy, Frere:1983ag, Derendinger:1983bz, 
Greene:1986th, Ellis:1986mq, Durand:1988rg, Drees:1988fc, Ellis:1988er, 
Pandita:1993tg, Pandita:1993hx, Ellwanger:1993xa, Ananthanarayan:1995zr,
Ananthanarayan:1995xq, Ananthanarayan:1996zv, Ellwanger:1996gw, Elliott:1994ht, 
King:1995vk}.  

% CMSSM

Over the last two decades, due to its simplicity and elegance, the constrained 
MSSM (CMSSM) and the minimal supergravity-motivated (mSuGra) model became a 
standard in supersymmetry phenomenology.  Guided by this, within the NMSSM, we 
impose the universality of sparticle masses, gaugino masses, and trilinear 
couplings at the grand unification theory (GUT) scale, thereby defining the
next-to-minimal supergravity-motivated (NmSuGra) model.
%
% CNMSSM
% 
This approach ensures that all dimensionful parameters of the NMSSM scalar 
potential also naturally arise from supersymmetry breaking in a minimal fashion.  
NmSuGra also reduces the electroweak and dark matter fine-tunings of mSuGra.

% CNMSSM likelihood analysis

Using a simple likelihood analysis, first we identify the parameter regions of the
NmSuGra model that are preferred by the present experimental limits from collider, 
astrophysical, and various low energy measurements.  We combine theoretical 
exclusions with limits from the CERN Large Electron-Positron (LEP) collider, the 
Fermilab Tevatron, NASA's Wilkinson Microwave Anisotropy Probe (WMAP) satellite 
(and other related astrophysical measurements), the Soudan Cryogenic Dark Matter 
Search (CDMS), the Brookhaven Muon g$-$2 Experiment, and various b-physics 
measurements including the rare decay branching fractions $b \to s \gamma$ and 
$B_s \to l^+ l^-$. 

% CNMSSM detectability

Next we show that the favored parameter space can be detected by a combination 
of the LHC and an upgraded CDMS, provided that the 3 $\sigma$ discrepancy 
between the experimental and the standard theoretical values of the anomalous 
magnetic moment of the muon ($\Delta a_{\mu}$) persists in the future. 
According to the latest calculations, the theoretical uncertainty of $\Delta 
a_{\mu}$ is now under control, which makes it a powerful discriminator at the 99 
\% confidence level (CL) \cite{Stockinger:2007pe, 
Passera:2008jk}\footnote{Recent studies show that additional isospin-breaking 
corrections lower the $\tau$-based determination of the hadronic leading-order 
contribution to $a_\mu^{SM}$, bringing it to good agreement with the $e^+e^-$ 
based determination \cite{Passera:2008jk}.}.  If, due to new experimental input, 
the $\Delta a_{\mu}$ discrepancy decreases, our results indicate that allowed 
regions of the NmSuGra model may not be reachable by either the LHC or a 1-ton 
equivalent of CDMS.  The detection of these regions will provide a new 
experimental challenge.

\section{The NmSuGra model}

% CNMSSM & cNMSSM

In this work, we adopt the superpotential 
\begin{eqnarray}
 W = W_{Y} + \lambda \hat{S} \hat{H}_u \cdot \hat{H}_d + \frac{\kappa}{3} \hat{S}^3,
\label{eq:W}
\end{eqnarray}
where $W_{Y}$ is the MSSM Yukawa superpotential \cite{Peskin:2008nw}, $\hat{S}$ 
($\hat{H}_{u,d}$) is a standard gauge singlet ($SU(2)_L$ doublet) chiral 
superfield, $\lambda$ and $\kappa$ are dimensionless couplings, and 
$\hat{H}_u \cdot \hat{H}_d = \epsilon_{\alpha\beta} \hat{H}_u^\alpha \hat{H}_d^\beta$
with the fully antisymmetric tensor normalized as $\epsilon_{11} = 1$.  
The corresponding soft supersymmetry breaking terms are
\begin{eqnarray}
 \mathcal{L}^{soft} = \mathcal{L}^{soft}_{MSSM} + m_S^2 |S|^2 + \nonumber \\
 (\lambda A_\lambda S H_u \cdot H_d + \frac{\kappa A_\kappa}{3} S^3 + h.c.),
\end{eqnarray}
where $\mathcal{L}^{soft}_{MSSM}$ contains the mass and Yukawa terms but not the 
$B \mu$ term.

% domain wall stabilization (hep-ph/9809475)

The superpotential (\ref{eq:W}) possesses a global $Z_3$ symmetry which is 
broken during the electroweak phase transition in the early universe.  The 
resulting domain walls should disappear before nucleosynthesis; however $Z_3$ 
breaking (via singlet tadpoles) leads to a vacuum expectation value (vev) for 
the singlet that is much larger than the electroweak scale.  Thus the 
requirement of the fast disappearance of the domain walls appears to destabilize 
the hierarchy of vevs in the NMSSM.  Fortunately, in 
Ref.~\cite{Panagiotakopoulos:1998yw, Panagiotakopoulos:2000wp} is was shown that, 
by imposing a $Z_2$ R-symmetry, both the domain wall and the stability problems 
can be eliminated.  Following \cite{Panagiotakopoulos:1998yw}, we assume that 
tadpoles are induced, but they are small and their effect on the phenomenology 
is negligible.

% CNMSSM

We also assume that the soft masses of the gauginos unify to $M_{1/2}$, those of 
the sfermions and Higgses to $M_0$, and all the trilinear couplings (including 
$A_\kappa$ and $A_\lambda$) to $A_0$ at the GUT scale.  This leaves nine free 
parameters: $M_0$, $M_{1/2}$, $A_0$, $\langle H_u \rangle$, $\langle H_d 
\rangle$, $\langle S \rangle$ (the Higgs and singlet vevs), $m_S$, $\lambda$ and 
$\kappa$. The three minimization equations for the Higgs potential 
\cite{Hugonie:2007vd} and $\langle H_u \rangle^2 + \langle H_d \rangle^2 = v^2$ 
(here $v = \sqrt{2} /g_2$ is the standard Higgs vev) eliminate four parameters. 
With the introduction of $\mu = \lambda \langle S \rangle$, and $\tan\beta = 
\langle H_u \rangle/\langle H_d \rangle$, our remaining free parameters are
\begin{eqnarray}
M_0, ~ M_{1/2}, ~ A_0, ~ \tan\beta, ~ \lambda, ~ {\rm sign}(\mu).
\label{eq:5Para}
\end{eqnarray}

% CNMSSM, others
 
Constrained versions of the NMSSM have been studied in the recent literature. 
The most constrained version is the cNMSSM \cite{Djouadi:2008yj} with $m_S = 
M_0$.  In other cases the $A_\kappa = A_\lambda$ relation is relaxed 
\cite{Hugonie:2007vd}, and/or $\kappa$ is taken as a free parameter 
\cite{Belanger:2005kh,Cerdeno:2007sn}, or the soft Higgs masses are allowed to 
deviate from $M_0$ \cite{Djouadi:2008uw} giving less constrained models.  In the 
spirit of the CMSSM/mSuGra, we adhere to universality and use only $\lambda$ to 
parametrize the singlet sector.  This way, we keep all the attractive features of 
the CMSSM/mSuGra while the minimal extension alleviates problems rooted in the 
MSSM.

% codes

Our goal is to show that the experimentally favored region of the NmSuGra model 
can be discovered by nascent experiments in the near future.  To this end, we 
use the publicly available computer code NMSPEC \cite{Ellwanger:2006rn} to 
calculate the spectrum of the superpartner masses and their physical couplings 
from the model parameters given in Eq.~(\ref{eq:5Para}).  Then, we use 
NMSSMTools 1.2.5, an extensively modified version of ISATools and that of 
DarkSUSY 4.1 \cite{Gondolo:2005we} to calculate the relic density of neutralinos 
($\Omega h^2$), the spin-independent neutralino-proton elastic scattering cross 
section ($\sigma_{SI}$), the NmSuGra contribution to the anomalous magnetic 
moment of the muon ($a_{\mu}^{NmSuGra}$), and various b-physics related 
quantities.  

% chi^2

For each set of the model parameters, we quantify the experimental preference in 
terms of
\begin{eqnarray}
\sqrt{\chi^2} = \biggl(\sum_{i=1}^7
\Bigl(\frac{m_i^{experiment}-m_i^{NmSuGra}}{\sigma_i}\Bigr)^2\biggr)^{1/2}
\label{eq:Chi2}
\end{eqnarray}
where $m_i$ is the central value of a physical quantity measured by an 
experiment or calculated in the NmSuGra model, and $\sigma_i$ is the combined 
experimental and (where available) theoretical uncertainty.  The sum includes 
the experimental upper limits \footnote{The WMAP limit on the abundance of dark 
matter, for example, is only used as an upper bound, that is the $\Omega h^2$ 
contribution to $\chi^2$ vanishes if the calculated relic density is below the 
experimental central value.} for 

(1) $\Omega h^2 = 0.1143 \pm 0.0034$ \cite{Komatsu:2008hk}, 

(2) $Br(B_s \to \mu^+\mu^-) = 5.8 \times 10^{-8}$ (95 \% CL) \cite{Barberio:2006bi}, and 

(3) $\sigma_{SI}$ by CDMS \cite{Ahmed:2008eu}, 

\noindent the LEP lower limits of the lightest scalar Higgs and chargino masses 
(which can be approximately stated as) \cite{Abbiendi:2003sc}:

(4) $m_h > 114.4$ GeV for $\tan\beta \stackrel{<}{\sim} 10$ 

~~~~~$m_h > 91$ GeV for $\tan\beta \stackrel{>}{\sim} 10$,

(5) $m_{\tilde{W}_1} > 104$ GeV,

\noindent and the central values of 

(6) $\Delta a_\mu = 29.5 \pm 8.8 \times 10^{-10}$ \cite{Stockinger:2007pe}, 

(7) $Br(b \to s \gamma) = 3.55 \pm 0.26 \times 10^{-4}$ \cite{Barberio:2006bi}.  

\noindent The related uncertainties are given above at 68 \% CL, unless stated otherwise.
Theoretical uncertainties are calculated using NMSSMTools for the b-physics 
related quantities.  Among the standard input parameters, $m_b(m_b) = 4.214$ GeV 
and $m_t^{pole}=171.4$ GeV are used.

\begin{figure}[t]
\includegraphics[width=0.49\textwidth,height=0.49\textwidth]{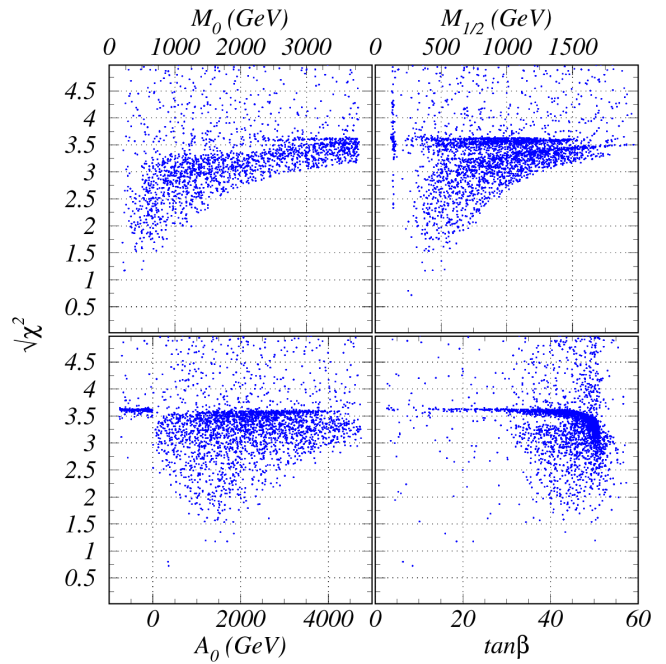}
\caption{\label{fig:Chi2VsInput} 
Square-root of the exponent of the likelihood function vs. four of the NmSuGra 
input parameters for a random sample of models.  The combination of the 
experimental quantities included in $\chi^2$ favor low values of $M_0$, 
$M_{1/2}$ and $|A_0|$.
}
\end{figure}

A glance at experimental likelihood reveals a significant statistical preference 
for relatively narrow intervals of $M_0$, $M_{1/2}$ and $A_0$. This is 
illustrated in FIG.~\ref{fig:Chi2VsInput}, which shows $\sqrt{\chi^2}$ as the 
function of $M_0$, $M_{1/2}$, $A_0$ and $\tan\beta$ for ${\rm sign}(\mu) > 0$ 
for a randomly selected set of models \footnote{Positive values of the $\mu$ 
parameter are favored by $\Delta a_\mu$ and $Br(b \to s \gamma)$ 
\cite{Barger:2005hb}.}.  From FIG.~\ref{fig:Chi2VsInput}, we conclude that it is 
enough to examine the low $M_0$, $M_{1/2}$ and $|A_0|$ region, since the rest is 
disfavored by the above combination of the experimental data at (or more than) 
99 \% CL. Similarly to the CMSSM/mSuGra, at high values of $M_0$, $M_{1/2}$ and 
$|A_0|$, $\chi^2$ is dominated by $\Delta a_\mu$.  

Based on the above, we limit our study to the following ranges of the continuous 
parameters \footnote{These limits arising mainly from the anomalous magnetic moment of 
the muon are very similar to the ones in the CMSSM \cite{Ellis:2007fu}.}:
\begin{eqnarray}
0 < M_0 < 4 ~{\rm TeV}, ~~ 0 < M_{1/2} < 2 ~{\rm TeV}, \nonumber \\
0 < |A_0| < 5 ~{\rm TeV}, ~~ 1 < \tan\beta < 60, ~~ 0.01 < \lambda < 0.7. 
\label{eq:5ParaRange}
\end{eqnarray}
The upper limit on $\lambda$ arises from the requirement that its running 
remains perturbative as it evolves up to the GUT scale.  Restricting $M_0$, 
$M_{1/2}$ and $A_0$ to such values also appears consistent with the electroweak 
precision data \cite{Ellis:2007fu} and greatly reduces fine tuning.

\section{Detectability of NmSuGra}

\begin{figure}[t]
\includegraphics[width=0.49\textwidth,height=0.49\textwidth]{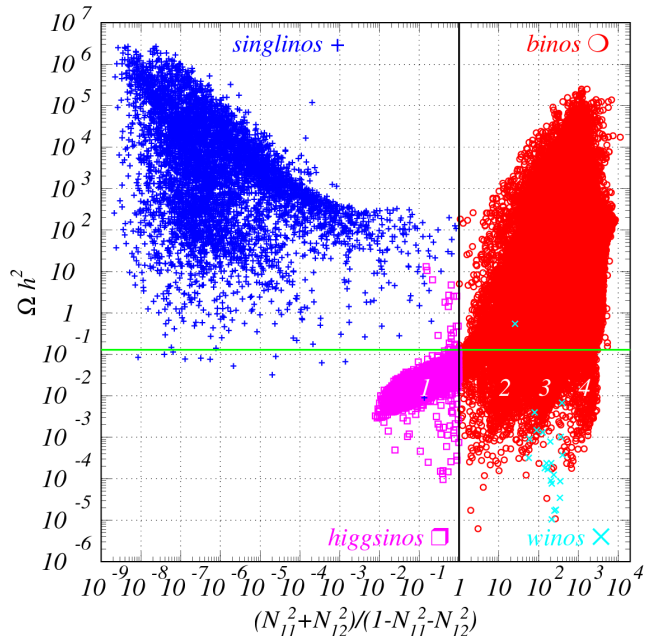}
\caption{\label{fig:Oh2vsAdMix} 
Relic abundance of the lightest neutralino as the function of its gaugino 
admixture.  Right of the vertical line the neutralino is mostly gaugino. The 
horizontal lines shows the WMAP upper limit (95 \% CL).
}
\end{figure}

Having defined the NmSuGra model and constraining the range of its parameters, 
we set out to show that this parameter region will be detectable by the LHC and 
an upgraded CDMS detector.  To this end, we randomly select about 20 million 
models in the range defined in Eq.~(\ref{eq:5ParaRange}), and for each model 
point evaluate Eq.~(\ref{eq:Chi2}).
Two million theoretically allowed representative model points are projected in 
FIG.~\ref{fig:Oh2vsAdMix} to the $\Omega h^2$ vs. $(N_{11}^2+N_{12}^2)/(1-
N_{11}^2-N_{12}^2)$ plane \cite{Barger:2007nv}.  This plane of $\Omega h^2$ and 
the neutralino mixing matrix elements ($N_{ij}$) is a good indicator of the 
gaugino, higgsino and singlino admixture of the lightest neutralino.  From 
FIG.~\ref{fig:Oh2vsAdMix} it is evident that the WMAP upper limit (green 
horizontal line) allows models with mostly bino- (red) and higgsino-like 
(magenta) lightest neutralino, while the fraction of allowed models with 
singlino-like (blue) dark matter is negligible.  As such the NmSuGra model is 
very similar to the CMSSM/mSuGra. 

% Oh^2 vs. AdMix

The similarity is even more evident when we examine the grouping of the WMAP 
allowed model points.  By checking mass relations and couplings, we can easily 
establish that the various branches (denoted by 1, 2, 3 and 4) belong to models 
with distinct neutralino (co)annihilation mechanisms well known from the 
CMSSM/mSuGra \cite{BalazsCarter2}.  Branch 4 contains only models with dominant 
neutralino-stop coannihilation, while branch 3 corresponds neutralino-stau 
coannihilation.  Branch 2 represents the Higgs resonance corridors, and branch 1 
is the equivalent of the CMSSM/mSuGra focus point region. 

% Oh^2 vs. AdMix detectability

\begin{figure}[t]
\includegraphics[width=0.49\textwidth,height=0.49\textwidth]{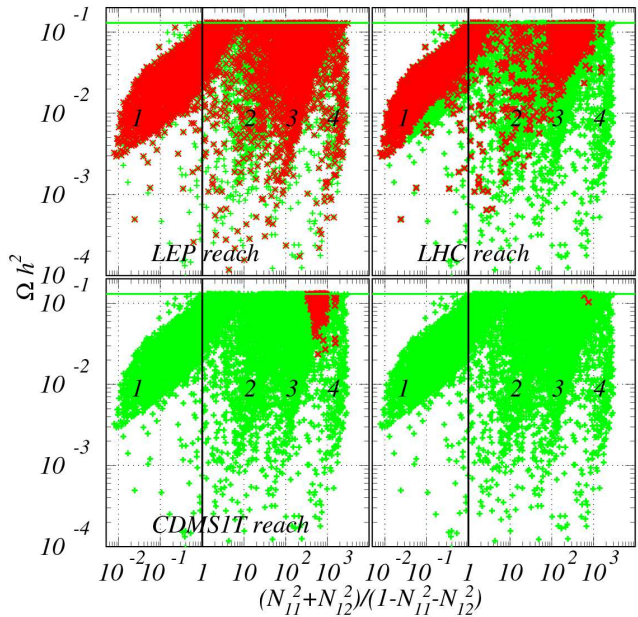}
\caption{\label{fig:Oh2vsAdMix5}
As FIG.~\ref{fig:Oh2vsAdMix} except green models can be reached by the combination 
of LEP, the LHC and CDMS1T, while red ones cannot.  The last frame dismisses
% experimental reach including g$-$2 (at the present 95 \% CL).
experimentally inaccessible points which have $\sqrt{\chi^2}>3$. 
}
\end{figure}

% LEP

To gauge the detectability of the NmSuGra model, first we identify model points 
that could have been seen at LEP.  We require the lighter chargino to be lighter 
than 103.5 GeV or the lightest scalar Higgs to be lighter than 114.4 GeV.  We 
relax the latter to $m_{H_{p_1}}$ when the mass of the lightest pseudo-scalar Higgs 
($m_{H_{p_1}}$) approaches that of the lightest scalar ($m_{H_{s_1}}$) 
\cite{LHWG200501}.  We do not apply the 114.4 GeV LEP Higgs limit when we 
encounter either a mostly singlet lightest Higgs or when $m_{H_{s_1}} < m_{H_{p_1}}$ 
\cite{Barger:2006dh,Djouadi:2008uw}.
Focusing our attention to the lower right corner of FIG.~\ref{fig:Oh2vsAdMix}, 
in the top left frame of FIG.~\ref{fig:Oh2vsAdMix5} we show the same model 
points in different coloring.  Models detectable by LEP are colored green, that 
is the green points represent the reach of LEP for the NmSuGra model.  The red 
colored models pass the above LEP constraints, i.e. are allowed by LEP.  Just as 
in the CMSSM/mSuGra the neutralino-stop coannihilation region is mostly covered 
by LEP.

% LHC reach explained

To estimate the LHC reach, we rely on the similarity between the mSuGra and 
NmSuGra models.  According to Ref.~\cite{Baer:2003wx} the reach of the LHC for 
mSuGra can be well approximated by the combined reach for gluinos and squarks. 
Based on this, if either the gluino mass is below 1.75 TeV, or the geometric 
mean of the stop masses is below 2 TeV for a given model point, we consider it 
discoverable at the LHC.  While this is an approximate statement that has to be 
supported by a detailed study of LHC event generation in the given models, in 
the light of Ref.~\cite{Baer:2003wx} it is a conservative estimate of the LHC 
reach.

% LHC reach

The top right frame of FIG.~\ref{fig:Oh2vsAdMix5} shows the model points that 
can be reached by LEP and the LHC using the above criteria. As in the 
CMSSM/mSuGra, most of the slepton coannihilation and the bulk of the Higgs 
resonance branches are covered by the LHC.  A good part of the focus point is 
also within reach of the LHC, with the exception of models with high $M_0$ 
and/or $M_{1/2}$.  

% CDMS reach

The bottom left frame of FIG.~\ref{fig:Oh2vsAdMix5} shows the reach of a one ton 
equivalent of CDMS (CDMS1T).  As expected from the CMSSM/mSuGra, the rest of the 
focus point and most of the remaining Higgs resonances are in the reach 
of CDMS1T.  The small number of models that remain inaccessible are all located 
in regions that have relatively low $M_0$ and high $M_{1/2}$ with dominant 
neutralino annihilation via s-channel Higgs resonances.  The NmSuGra contribution 
to $\Delta a_\mu$ in these model points is outside the preferred 99 \% CL 
region as shown by the last frame.

% Singlets
%
While we focused our discussion and plotting on the higgsino and gaugino 
regions, we carefully checked that LEP, the LHC and the upgraded CDMS 
experiments combined can also detect models with WMAP allowed singlino-like 
dark matter.  
The reason for this is the following.  The singlino-like lightest neutralinos 
typically take part in more than one (co)annihilation mechanisms to satisfy the 
WMAP limit.  While being on a Higgs resonance, they also coannihilate with 
sfermions.  In this case, either being in the sfermion coannihilation region 
ensures the LHC detectability or the Higgs resonance is strong enough to enhance 
recoil detection.

Furthermore, the universal mass relations ensure that the lightest scalar Higgs 
never decays to a pair of the pseudo-scalar Higgs bosons in the 
phenomenologically allowed region of NmSuGra, just as in the cNMSSM 
\cite{Djouadi:2008yj}.  Similarly, the lightest singlinos, allowed by the 
present experimental constraints are always heavier than about 100 GeV.  
Assuming that the NmSuGra contribution to the anomalous magnetic moment of the 
muon is larger than a minute $3.1 \times 10^{-10}$ constrains slepton and 
chargino masses below 3 and 2.5 TeV, respectively.  Since universality restricts 
the mass hierarchy within NmSuGra, the resulting mass spectrum is typically 
mSuGra-like.  Thus, the cascade decays and their signatures at LHC are not 
expected to be significantly deviate from that of the mSuGra case.  The most 
typical NmSuGra decay cascade at the LHC would be gluino $\to$ squark, quark 
$\to$ chargino/neutralino, W/Z $\to$ neutralino, SM particles.

\section{Conclusions}

Analyzing the next-to-minimal supergravity motivated (NmSuGra) model, we found 
that the LHC and an upgraded CDMS experiment will be able to discover the 
experimentally favored region of this model, provided that the present deviation 
between the experimental and standard theoretical values of the muon anomalous 
magnetic moment prevails.  If, due to future experimental input, the $g-2$ 
constraint weakens, then certain parameter regions of the NmSuGra model will 
have to be detected in alternative ways.

\begin{acknowledgments}

We thank M. Carena, U. Ellwanger, A. Menon, D. Morrissey, C. Munoz and C. Wagner 
for invaluable discussions on various aspects of the NMSSM.  This research was 
funded in part by the Australian Research Council under Project ID DP0877916.

\end{acknowledgments}

% \bibliography{references}

\end{document}